# The New SI and the CODATA recommended values of the fundamental constants 2017 compared with 2014, with a Comment to "Possolo *et al.*, Metrologia 55 (2018) 29"

Franco Pavese

*formerly National Research Council, Istituto di Metrologia "G.Colonnetti" (from 2006, Istituto Nazionale di Ricerca Metrologica), Torino, Italy*

e-mail: frpavese@gmail.com, frpavese@yahoo.com

This note comments on the special CODATA 2017 adjustment of the fundamental constants of July 2017 involved in the revision of the SI (based on: P.J. Mohr *et al.*, Data and Analysis for the CODATA 2017 Special Fundamental Constants Adjustment for the Revision of the SI, Metrologia 2018, September 2017 preprint), with a comparison to the CODATA 2014 adjustment. In Appendix, a comment is also added on the paper Possolo *et al.*, Metrologia 55 (2018) 29 on the Planck constant. The previous versions of this manuscript (also available on ArXiv) were aiming at pointing out some standing features of the present and future CODATA method in the light of the CODATA table of 2014 recommended values for the fundamental constants. A comprehensive discussion on this and related issues is becoming very important in view of the foreseen revision of the International System (SI) of measurement units in 2018. The present features may still raise doubts on a possible mixing of physical reasons of general validity in science with some needs specific of legal aspects of metrology concerning the SI. This illustration was adjourned in version 3 (v3) according to the CCU 2016 draft of the 9[th] SI Brochure and to the outcomes of its 22[th] meeting, with a note about the intended need of a 8[th] digit in the stipulated value of $k$.

**Introduction**

In 2015 the CODATA Group on fundamental constants published a summary of the results of their 2014 adjustment of the numerical values of the fundamental constants, [Mohr *et al.* 2015] following the full publication in 2012 of the 2010 adjustment [Mohr *et al.* 2012]. In the mean time, the Conférence Générale des Poids et Measures (CGPM) examined in 2011 and 2014 the proposal presented by the Bureau International des Poids et Mesures (BIPM) to deeply revise the definition of the International System of Units (SI) by using physical constants for its definition. [BIPM 2016] As a long since metrologist, for some years I am observing this process, and in particular the CODATA work on numerical outcomes, obtained with the application of the Least Squares Adjustment (LSA) method, which are normally used in the official BIPM documents as the ones to be used to define the numerical values of those constants in the revised SI definition. In early 2014 a Letter devoted to this CODATA issue was published [Pavese 2014] with a Reply to it [de Mirandes 2014] on the same Journal.

Then, in 2014 [Mohr *et al.* 2015] there were changes in the numerical *values* of the five constants that are relevant to the present proposal concerning the SI: they were used in the 2016 CCU Draft of the SI Brochure [CCU 2016]. In applying its LSA method, the CODATA still kept fixed not only $\mu_0$, but also $c$, and thus also $\varepsilon_0$. Subsequent to the CCU 22[th] meeting, we learned from the minutes that the next CODATA adjustment would have been the basis for the final proposal to the CGPM in the fall of 2008. Therefore, the new constants' adjusted values will be published in 2018, with a preprint available since September 2017, with also a pre-stated number of digits without attached uncertainty.

Earlier in 2017, a request to stipulate the numerical value of $k$ with 8 digits was commented in v3, justified by the need to fulfil Resolution 1 of the 24th CGPM (2011) [CGPM 2011] implying that "a sufficient number of digits for each constant is required such that the numerical value of each constant m($K$), $\mu_0$, $T_{TPW}$, and $M(^{12}C)$ expressed in the 'old' and 'new' SI units should be strictly equal, i.e., *consistent with their standard uncertainties*, at the time of the redefinition" (emphasis added) [CCT 2017]. That point was previously already raised at the 2016 CCU 22[th] meeting [CCU 22[th]].

The present note continues to track the evolution of the work within CCU and CODATA, now after the issue of the 2017 CODATA special adjustment, as contained in the distributed preprint of the article submitted to Metrologia, and after the CCU 2017 Recommendation to CIPM, outlining some current possible consequences of it.

**The CODATA 2017 *special* adjustment**

The 2017 adjustment ([Mohr *et al.* 2017, Newell *et al.* 2017], to be published in 2018, see References) is called by CODATA "special", because "as also noted by Newell *et al.* (2018) (*still to appear at this date*], the procedures used for the 2017 Special Adjustment are the same as those used for the 2014 regularly scheduled CODATA adjustment and its 1998, 2002, 2006, and 2010 predecessors", but "the purpose of the CODATA 2017 Special Adjustment is to obtain best numerical values for *h, e, k,* and $N_A$, not to provide a complete set of recommended values to replace the 2014 set" and thus "to maintain the continuity of CODATA adjustments and to aid understanding, the starting point of the 2017 Special Adjustment is the input data included in the final adjustment on which the 2014 CODATA recommended values are based".
From it, we understand that "the 22 input data omitted from that final adjustment because of their low weight are not considered for inclusion in the 2017 Special Adjustment".
This means that both 2014 and 2017 adjustments were not performed using *all* the constants used in the previous ones and until 2010. This is a basic and delicate issue, since it is known that the LSA method puts a relationship between *all* the constants involved. [Taylor *et al.* 1969] There is no published evidence available that the last adjustments can be considered equivalent, except the CODATA statement "because of their [*of the omitted constants*] low weight", not a quantitative one, so one can only trust CODATA about the correctness of their work.

**Illustration of the observed problems**

*Changes in the recommended values of the constants*
In paper [Mohr *et al.* 2017], one can observe the changes of the adjusted values of the relevant constants with respect to the 2014 ones. The paper, as usual, only discusses the CODATA uncertainties, not the changes in value (the "adjustment"). However, for the New SI, where the numerical values are supposed to be then stipulated, also the values being the 'best' in a strict statistical sense are an important issue. Table 1 and Fig. 1 report a comparison of the 2006–2017 data (*c* is already fixed). Also $m_u$ is included, considered in [Fletcher *et al.* 2015].
The trend of change of *e*, *h* and $N_A$ (and $m_u$) does not indicate that one can be confident that the 2017 ones might represent the last significant variation. All changes in the adjusted values (see column "Change" in Table 1) were larger than the final uncertainty, except for *k*.
On the other hand, in Fig. 1 the 2017 values for *e*, *h* and $N_A$ (and $m_u$) are clearly not "metrologically compatible" [VIM] (for a 1s level) with the 2006 value.

Table 1. Change in numerical value of the CODATA adjustments **2006-2017** for the constants involved in the New SI definition. [ Mohr *et al.* 2015, Newell *et al.* 2017, Fletcher *et al.* 2015]

| Constant | CODATA | Numerical value * | s.d./relative ×$10^7$ | Change | Total shift |
|---|---|---|---|---|---|
| $e \times 10^{19}$ | 2006 | 1.602 176 4$_9$ | 0.4/0.23 | – | |
| | 2010 | 1.602 176 5$_7$ | 0.35/0.19 | *8 ×10$^{-8}$* | |
| | 2014 | 1.602 176 6$_2$ | 0.1/0.06 | *5 ×10$^{-8}$* | 1.3 ×10$^{-7}$ |
| | 2017 | 1.602 176 634 [a] | 0.08/0.05 | *1.4 ×10$^{-8}$* | 1.4$_4$ ×10$^{-7}$ |
| $h \times 10^{34}$ | 2006 | 6.626 069$_0$ | 3.3/0.50 | – | |
| | 2010 | 6.626 069$_6$ | 2.9/0.44 | *6 ×10$^{-7}$* | |
| | 2014 | 6.626 070 04 [b] | 0.8/0.12 | *4.4 ×10$^{-7}$* | 10.4 ×10$^{-7}$ |
| | 2017 | 6.626 070 1$_5$ [b] | 0.7/0.11 | *1.1 ×10$^{-7}$* | 11.5 ×10$^{-7}$ |
| $N_A \times 10^{-23}$ | 2006 | 6.022 141$_8$ | 3/0.50 | – | |
| | 2010 | 6.022 141$_3$ | 2.7/0.45 | *−5 ×10$^{-7}$* | |
| | 2014 | 6.022 140$_8$ [c] | 0.7/0.12 | *−4 ×10$^{-7}$* | −9 ×10$^{-7}$ |
| | 2017 | 6.022 140 7$_6$ [c] | 0.6/0.10 | *+1 ×10$^{-7}$* | −8 ×10$^{-7}$ |
| $k \times 10^{23}$ | 2006 | 1.380 65$_{04}$ [d] | 24/17 | – | |
| | 2010 | 1.380 64$_{88}$ [d] | 13/9.4 | *−1.6 ×10$^{-6}$* | |
| | 2014 | 1.380 648$_5$ | 8/5.8 | *−0.3 ×10$^{-6}$* | −1.9 ×10$^{-6}$ |
| | 2017 | 1.380 649 [e] | 5/3.6 | *+0.5 ×10$^{-6}$* | −1.4 ×10$^{-6}$ |
| $m_u \times 10^{27}$ | 2006 | 1.660 538 7$_8$ | 0.8/0.48 | – | |
| | 2010 | 1.660 538 9$_2$ | 0.7/0.42 | *12 ×10$^{-8}$* | |
| | 2014 | 1.660 539 04 | 0.2/0.12 | *14 ×10$^{-8}$* | 2.6 ×10$^{-7}$ |
| | 2017 | — [f] | | | |

(please note various *errata-corrige* with respect to v4 of this manuscript, also reflecting on Fig. 1).
* The smaller-case digits are *uncertain* (in 2017 *rounded according to* [Mohr *et al.* 2017]) taken from the CODATA two-digit format, except for *k*—see note c) and d). Changes exceeding the uncertainty are shown in *italics*.
[a] The 2017 CODATA outcome is 1.602 176 6341(83), therefore the numerical value can be as low as 1.602 176 6258 and as high as 1.602 176 6424, thus involving also the preceding digit.
[b] The 2014 CODATA outcome is 6.626 070 040(81), therefore the numerical value can be as low as 6.626 069 959, thus involving also the preceding digit. Similarly for the 2017 one, the CODATA outcome is 6.626 070 150(69).
[c] The 2014 CODATA outcome is 6.022 140 857(74), therefore the numerical value can be between 6.022 140 783 and 6.022 140 931, thus involving the preceding digit. Similarly for the 2017 one, the CODATA outcome is 6.022 140 758(62).
[d] Two digits are shown because the rounding involves also the preceding digit.
[e] The CODATA outcome is 1.380 649 03(51), thus the rounding does not include uncertain digits (the only occurrence in the Table). However, the numerical value can be as low as 1.380 648 50, thus involving the last digit, 9.
[f] Not published yet.

In Table 2 the same analysis of Table 1 is performed, but for constants *linked* to the constants of Table 1. The numerical value is the adjusted one taken from [Mohr *et al.* 2017] and previous CODATA reports. Again, most changes in the adjusted values (see column "Change" in Table 1) were larger than the final uncertainty and the 2016 values are not "metrologically compatible" [VIM] (for a 1s level) with the 2006 value, except for *R* in 2014. Asymptotic values do not look to be reached yet for *R* and $K_J$.

Table 2. Change in numerical value of the CODATA adjustments **2006-2017** for derived constants. [Mohr *et al.* 2015, Mohr *et al.* 2017]

| Constant | CODATA | Numerical value * | s.d./relative ×10$^7$ | Change | Total shift |
|---|---|---|---|---|---|
| $\alpha \times 10^3$ | 2006 | 7.297 352 53$_8$ | 0.050/0.007 | – | |
| | 2010 | 7.297 352 57$_0$ [a] | 0.024/0.003 | *+32 ×10$^{-9}$* | |
| | 2014 | 7.297 352 56$_6$ | 0.017/0.002 | –4.0 ×10$^{-9}$ | *28 ×10$^{-9}$* |
| | 2017 | 7.297 352 56$_5$ [c] | 0.0023/0.0003 | –1.0 ×10$^{-9}$ | –5 ×10$^{-9}$ |
| $R \times 10^6$ | 2006 | 8.314 47 | 150/18 | – | |
| | 2010 | 8.314 46$_2$ | 75/9 | *–8.0 ×10$^{-6}$* | |
| | 2014 | 8.314 46$_0$ [a] | 48/6 | –2.0 ×10$^{-6}$ | *–10 ×10$^{-6}$* |
| | 2017 | 8.314 47$_2$ [d] | 40/5 | *+12 ×10$^{-6}$* | *14 ×10$^{-6}$* |
| $K_J \times 10^{14}$ | 2006 | 4.835 978$_9$ | 1.2 | – | |
| | 2010 | 4.835 978$_7$ | 1.1 | –2.0 ×10$^{-7}$ | |
| | 2014 | 4.835 978 5$_3$ | 0.3 | *–1.7 ×10$^{-7}$* | *–3.7 ×10$^{-7}$* |
| | 2017 | — [b] | | | |
| $R_K \times 10^4$ | 2006 | 2.581 280 75$_6$ | 0.018 | – | |
| | 2010 | 2.581 280 74$_3$ | 0.008 | *–1.2 ×10$^{-8}$* | |
| | 2014 | 2.581 280 74$_{56}$ | 0.006 | +0.1 ×10$^{-8}$ | –1.1 ×10$^{-8}$ |
| | 2017 | — [b] | | | |

\* The smaller-case digits are *uncertain* (rounded to one from the CODATA two-digit format). CODATA adjusted values are shown. Changes exceeding the uncertainty are shown in *italics*.
[b] Two digits are shown because the rounding involves also the preceding digit.
[a] Conventional value used.
[c] In Mohr *et al.* 2017, the value of $\alpha$ can be obtained from the 2017 values of $h/e^2$ equal to $\mu_0 c/2\alpha$:
$\alpha = \mu_0 c\, e^2/2h$ = 7.297 352 565 10, where $\mu_0 c$ is stipulated. In addition, from two experiments:
7.297 352 565 77, and 7.297 352 574 02 (mean 7.297 352 567).
[d] From Table 1, computed as $R = kN_A$.

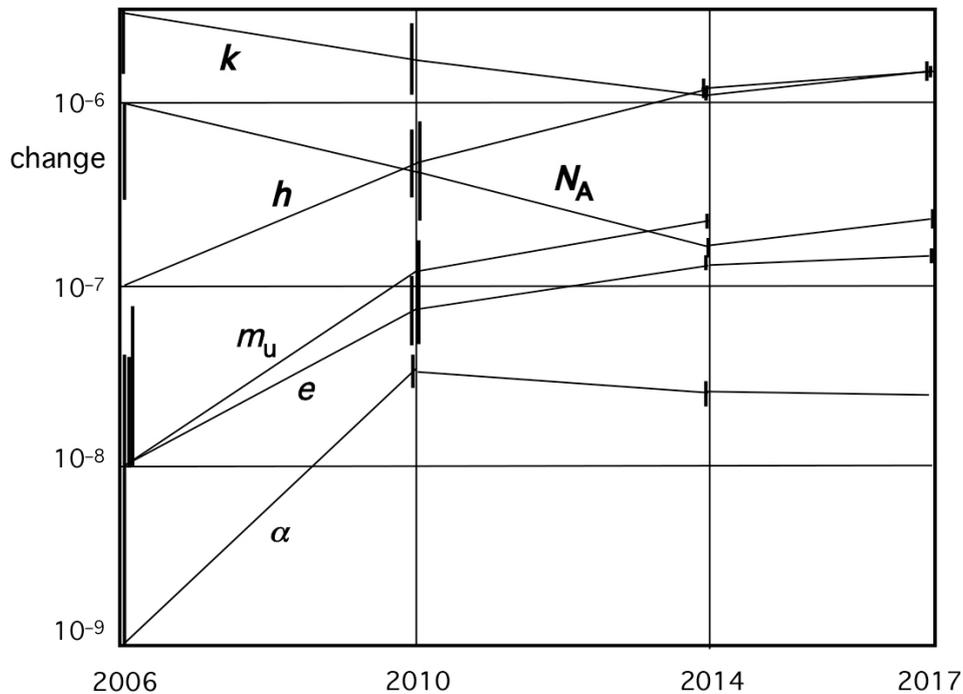

Fig. 1. Behaviour with time of the CODATA changes in adjusted values of some constants, with uncertainty bars. [Mohr *et al.* 2015, Mohr *et al.* 2017, Fletcher *et al.* 2015]

In Table 3 the adjusted values of the constants of Table 2 are compared with the corresponding results of their *computation* from other constants. One can notice the large differences in the level of agreement between the pairs of values, level that may be called the 'degree of consistency' for that group of constants. The 2010 differences in values for $\alpha$ and $R_K$ exceed the uncertainty reported in Table 2. Finally, the calculated uncertainties for $R$, as arising from the computation $R = kN_A$, are: $u \cdot 10^7 = 145, 78, 48, 40$ respectively, to be compared with the CODATA uncertainties reported in Table 2.

On the other hand, already for all 2014 data the degree of consistency is *the highest ever achieved*. That consistency is a basic requirement for the New SI, but one may wonder if it arises straight from the application of the LSA method, or is due to any further culling of the data—including by CODATA [Mohr et al. 2017].

In fact, this degree of consistency cannot arise simply from the fact that the defining equations reported in column 1 of Table 3 are included among the constraints of the LSA method, because, due to the "*intimate relationship*" set by the LSA method between *all* constants, recalled in [Taylor et al. 1969], there are *no reasons* for expecting a general increase of the degree of consistency as a mere consequence of a lowering of the uncertainties of the experimental data.

Table 3. Change in adjusted and computed values from other constants of the numerical values for the CODATA adjustments **2006-2017**.

| Constant | Year | Adjusted | Computed | Difference |
|---|---|---|---|---|
| $\alpha = \mu_0 c$ $e^2/2h$ $\times 10^3$ | 2006 | 7.297 352 5376 | 7.297 352 5368 | $8.0\ 10^{-13}$ |
| | 2010 | 7.297 352 5698 | 7.297 352 5755 | $-57.2\ 10^{-12}$ |
| | 2014 | 7.297 352 5664 | 7.297 352 5662 | $1.9\ 10^{-13}$ |
| | 2017 | — | 7.297.352 5651 | — |
| $R = kN_A$ | 2006 | 8.314 472 | 8.314 4725 | $-5.0\ 10^{-7}$ |
| | 2010 | 8.314 4621 | 8.314 4621 | 0.0 |
| | 2014 | 8.314 4598 | 8.314 4599 | $-1.0\ 10^{-7}$ |
| | 2017 | — | 8.314 4721 | — |
| $K_J = 2e/h$ $\times 10^{14}$ [a] | 2006 | 4.835 9789 | 4.835 978 909 | $1.0\ 10^{-9}$ |
| | 2010 | 4.835 9787 | 4.835 978 699 | $1.0\ 10^{-9}$ |
| | 2014 | 4.835 978 53 | 4.835 978 525 | 0.0 |
| | 2017 | — | — | — |
| $R_K = h/e^2$ $= \mu_0 c/2\alpha$ $\times 10^4$ [b] | 2006 | 2.581 280 756 | 2.581 280 7560 | $-3.0\ 10^{-10}$ |
| | 2010 | 2.581 280 7443 | 2.581 280 742 32 | $2.0\ 10^{-9}$ |
| | 2014 | 2.581 280 7456 | 2.581 280 745 61 | $-6.0\ 10^{-11}$ |
| | 2017 | — | — | — |

[a] Present conventional numerical value: $4.835\ 979 \times 10^{14}$ exact. [b] Present conventional numerical value: $2.581\ 2807 \times 10^4$ exact.

Until 2014 one was concerned about some of the actions stated in [CCU 22th] on the way to obtain the stipulated numerical numbers for the relevant constants. Then the CCU decided to mandate to CODATA the format of the stipulated numbers according to the so-called "case 3" model, by fixing 7 overall digits for all the 2017 values in the above Table. This is reported now in [Newell et al. 2017], where the number of digits for $e$, $h$, and $N_A$ is larger (the small case digits are affected by CODATA uncertainty):

$\{e\} \times 10^{19} = 1.602\ 176\ 6_{34}$,
$\{h\} \times 10^{34} = 6.626\ 070\ _{15}$,
$\{N_A\} \times 10^{-23} = 6.022\ 140\ _{76}$,
$\{k\} \times 10^{23} = 1.380\ 649$ (the next digit, not indicated, is a 0).

With the indicated rounding, all of them do *not* exclude, except $k$, the digit(s) affected by the CODATA uncertainty, contrarily to what was advised since [Pavese 2013].

The rational is that, the previous stipulated values should *remain unaltered if the "continuity principle"* has to be respected [CCU 2016], within the experimental uncertainty assessed by the experimental database—not within the CODATA database where the uncertainty is often lowered by the LSA treatment. In [Mohr *et al.* 2017], these "consistency factors" are listed as follows:

$[m(K)/(kg)_{rev}]/1 = 1.000000001(10)$ $[1.2 \times 10^{-8}]$    (i.e., ± 12(10) µg)
$[\mu_0/(Hm^{-1})_{rev}]/(4\pi \times 10^{-7}) = 1.00000000020(23)$ $[2.3 \times 10^{-10}]$
$[T_{TPW}/(K)_{rev}]/273.16 = 1.00000001(37)$ $[5.7 \times 10^{-7}]$    (i.e., ± 27 µK)
$[M(^{12}C)/(kgmol^{-1})_{rev}]/0.012 = 1.00000000037(45)$ $[4.5 \times 10^{-10}]$.

In general, for the constants that *will become adjustable*, namely $\mu_0$, thus also $\varepsilon_0$, with the application of the LSA method, the previous fixed values *will be preserved only within the above relative uncertainty*.

From the above, the 8$^{th}$, 9$^{th}$ or 10$^{th}$ digits are not justified by the 2017 experimental uncertainties of the $e$ ($u \approx 8 \times 10^{-9}$), $h$ ($u \approx 7 \times 10^{-8}$) and $N_A$ ($u \approx 6 \times 10^{-8}$), affecting the last fixed digit— nor would be for $k$ ($u \approx 5 \times 10^{-7}$).

*Comparison of the 2014 and 2017 CODATA adjustments*
It is significant to compare the 2014 and 2017 CODATA adjustments, where the list of the data compared in the following is reported in [Mohr *et al.* 2015] and [Mohr *et al.* 2017]. Here, this exercise is done for the Planck constant, also in support to the Appendix.
Figure 2 shows the original data and their uncertainties.

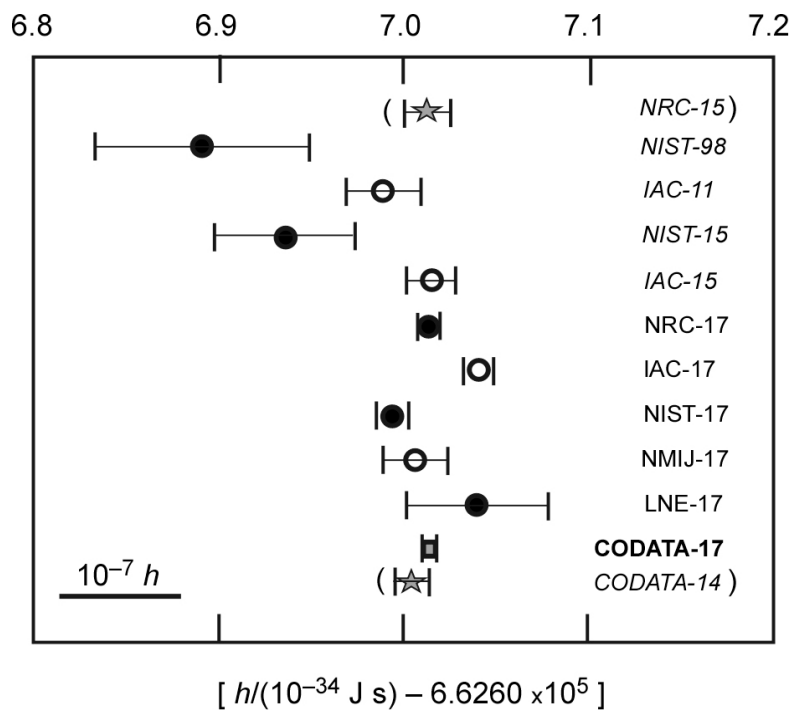

Fig. 2. Database used in 2014 (*italics*) [Mohr *et al.* 2015] and 2017 [Mohr *et al.* 2017] for the results on the Planck constant: open circle = using Kibble balance; black circles = using x-ray-crystal-density; star = included only in 2014. Uncertainty bars: $k = 1$. NIST-98 *omitted* in [Newell *et al.* 2017].

Clearly, apart NRC-15, the 2014 adjustment used a *subset* of the one available in 2017. Both were of the same type: a "special" adjustment dropping a number of constants with "ininfluent" correlation—as already reported hereinbefore. Previous and further determinations with larger uncertainties were omitted.

One has to note that *the number of data inconsistent with each other has increased* from 2014 to 2017, *while the uncertainty* associated to the result of the adjustments (CODATA recommended value) *has quite lowered*.

That looks to be a feature of the LSA method: it aims at checking the *consistency* of the results, and, for doing that, the original experimental *values* are "adjusted", i.e. *changed* in order to optimise the *final variance of the set*. In the case of inconsistent results, their associated uncertainties were increased in the CODATA analysis, as reported in the next Section.

However, more important, in those cases *the changes of the original values* (the "adjustment") are necessarily *characterised by being larger* than the respective *original* variances: unfortunately, this important feature is invisible and we cannot check it, since the values of the single adjustments are not provided in the CODATA Reports.

*Meaning and way to use the LSA method*

The LSA method needs several assumptions, which are now recalled in [Mohr *et al.* 2017] and, probably, in [Newell *et al.* 2018]. In particular, it is indicated that the initial "input data" were 138 and the relationships between them (the constraints) were 74; thus the degrees of freedom were 64. The Birge ratio was 1.07. They required some manipulation, since some weighted residuals $r_i = (x_i - \langle x_i \rangle)/u_i$— where $\langle x_i \rangle$ are the adjusted values—were unacceptably large: these numerical adjustments are nt available in the CODATA Reports. For them the uncertainties were expanded, so reducing their influence in the process. Kibble balance data (relevant to mass unit) were "particularly problematic". Finally, data having a limited influence on the final results were omitted. The final set included 134 data, 74 constraints and 60 degrees of freedom, with a Birge ratio 0.85.

Clearly, the results of the LSA method are *conditional* to the set of conditions to which the data were subjected, including the constants that, from the initial fixed $\mu_0$ extended to a fixed $c$, and thus to a fixed $\varepsilon_0$. After 2018, they will extend to further stipulated constants. No published study on LSA has ever provided a quantitative estimate of the effect of changed constraints, in particular when arising from non-adjusted constants.

When having fixed members of the set and other conditions 'filling-up' the degree of freedom, the conditional effect induced by their choice may be irrelevant in many other fields where the goal is only to check for the best consistency of the set. In the case of the use of the constants in measurement units, instead, also the *continuity* (in magnitude) of the base units through the change of the unit definitions is a must, so that the adjusted *values* should be considered to also optimise that continuity. This may place limitations to the allowable constraints. Also a study on this issue, specific of metrology, is not yet available, except by using the "consistency factors" shown before.

Long since it is a well-known fact that the LSA method causes an "*intimate relationship which exist* [is set by the LSA] *among least-squares adjusted values of the fundamental constants*", so that "*a significant shift in the numerical value of one will generally cause significant shifts in others*" [Taylor *et al.* 1969; Langenberg, Taylor 1970]. A published study on the changes in the adjustments posterior to 1983, in respect to leaving $c$, and thus also $\varepsilon_0$, adjustable or not, is not available. Also a study on the quantitative effect of having *limited the set of constants*, which may changes, in principle, the adjustement comparability since 2014 with previous adjustments, has not been published.

Another clear feature of the LSA method is that it optimises the set of data with respect to their *internal consistency*. This was occasionally very clear by comparing the resulting uncertainties with the actual original experimental uncertainties, where the uncertainty of the former is generally higher than that of the latter. Also in the 2017 adjustment, there is evidence of this effect, where the CODATA associated uncertainty stands at the lower bound of the interval of the experimental ones. For example, the final relative uncertainty associated with:

$h \times 10^{34}$ is $u_{rel} = 1 \times 10^{-8}$, while, for the 5 direct measurements, is $u_{rel} = 1-8 \times 10^{-8}$
$R \times 10^{6}$ (most used to measure $k$) is $u_{rel} = 5 \times 10^{-7}$ while, for the 10 direct measurements, is $u_{rel} = 3 \times 10^{-7}$ to $6 \times 10^{-6}$
$N_A \times 10^{-23}$ is $u_{rel} = 1 \times 10^{-8}$, while, for 4 the direct measurements, is $u_{rel} = 1-3 \times 10^{-8}$

Finally, a number of papers in the last decade provided elaborations using methods different from the LSA to obtain "best" values of these constants, e.g., [Bodnar *et al.* 2015, Bodnar *et al.* 2016, Mana 2016]. The resulting values are different from those obtained by CODATA. For example, for the Planck constant, $\{h\}$, based on 2010 data:

[Mana 2016]           6.626 070 073(94)
[Bodnar *et al.* 2016] 6.626 069 39(47)
[CODATA 2010]        6.626 069 58(23) —on which the above analyses are based
([CODATA 2017]       6.626 070 15 after stipulation)

This suggests, again, for the stipulation of an exact value of those constants, that, according to the precaution principle, CCU should drop from the 2017 CODATA-suggested exact values at least the last digit, which is affected also by CODATA original uncertainty, except for $k$.

The previous issues are basically 'only' metrologically relevant.
On the other hand, however, the CODATA analyses are of the greatest importance in a much wider context. Therefore, a too large influence of metrological constraints into the normal decisions of CODATA in the application of the LSA method (wider in usefulness than it is for the metrological issues) and in the resulting adjustments, might not be seen favourably by scientists in general.
The problem of LSA constraints was raised at the CCU 22[th] meeting. Apparently, the problem was ignored and there were even statements that all the constants are independent with each other, which is not true—see above.

In this respect, what happened in 1973 is exemplary of something that should not have happened. As reported in [Pavese 2014] "The 1973 adjustment preserved the value $c$ = 299 792 458 m s$^{-1}$ (*'Without intending to prejudge any future redefinition of the metre or the second, the CCDM suggested that any such redefinitions should attempt to retain this value provided that the data upon which it is based are not subsequently proved to be in error'* [9]), but not the uncertainty, set to 1.2 m s$^{-1}$ (4×10$^{-9}$ relative). The value, and its uncertainty, used instead for the 1973 LSA, as said in [Cohen, Taylor 1973], was the Evenson's one [Evenson *et al.* 1972], (299 792 456.2 ± 1.1) m s$^{-1}$ (3.5×10$^{-9}$ relative). Notice than [Mulligan 1976] reports for the same Evenson work the value (299 792 457.4 ± 1.1) m s$^{-1}$. Later, [Blaney *et al.* 1974] obtained 299 792 459.0(0.8) m s$^{-1}$"; apparently, this latter determination was not considered at all in the 1983 stipulation of $c$.

The constraint placed by the stipulations may become more influent *for science* after stipulation of more constants, as needed to the definition of the revised SI, which would not only make fixed the value of the constants appearing in the definitions, but will also make fixed the numerical value of others functionally depending on them (e.g. $R = kN_A$; others will also be constrained: e.g., $\alpha$ will then only depend on $\mu_0$ being then adjustable—but initially "consistent" with the previous exact value).
In all instances, the change of the constraints to the set does *automatically*, *in itself*, change more or less all values of the adjustable constants, irrespective to other reasons for adjustment.

These issues should be clarified in a more convincing way. Science, in general, should not be tied to these constraints.

**Conclusions**

In the last decade an extraordinary scientific effort, theoretical and experimental, has been performed for measuring a number of fundamental constants on stronger bases and with decreased uncertainty, bringing to extraordinary good results. However, one should never forget a risk that past experience has shown to be less than unexpected, and that is considered the usual state of affairs by reputed scientists working on human factors [Henrion, Fischhoff 2013]: the 'attraction' (also called "bandwagon" effect) of a target value or threshold associated uncertainty value.
In this situation, special provisions should be set to increase confidence that this occurrence is prevented, and that the present outcomes, namely the persistence in time of the stipulated values, can be checked in future.

The present case of the Kibble balance for $h$ (noted by CODATA and CCU) is paradigmatic: a stipulation in autumn 2018 will mean leaving the *status quo* of the 2017 CODATA adjustment, with the noted limitations for $h$ and the CODATA-suggested stipulated value.
In this respect, in [CCU 23$^{th}$], Recommendation U1 (2017), issued as a final recommendation in September 2017, the CCU was aware that "work is underway in NMIs to understand the cause for the dispersion of the experimental determinations of the Plank and Avogadro constants", as found by CODATA in the 2017 special adjustment, and as shown in Fig. 1, where the trend of the adjusted values from 2006 on does not seem to have reached a sufficient stability. However, the CCU concludes instead that "the numerical values [of the constants] provided by the CODATA … provide a sufficient foundation to support the redefinition" and recommends that "CIPM undertakes the necessary steps to proceed with the planned redefinition of [the units] at the 26$^{th}$ CGPM in 2018".
However, NMIs where non-consistent values were obtained would be obliged, since 2019, to use a correction for the results outcoming from their equipments to be used in the definition of their standards. On the other hand, a delay of a couple of years could instead resolve that major issue.

This dilemma can instead be solved by recognising, as suggested since some years by [Pavese 2017], that " … immediately after the change of definition, they (*the present standards*] still ensure the consistency of the old with the new units. This means respecting the ''principle of continuity'', within the uncertainties associated to the results obtained with the present-SI. It is an intrinsic property of the previous standards … for example, for an unforeseeable number of years also the copies of the IPK, and the IPK itself, if verified to remain stable in time, can still carry the present level in the traceability chain and consequently drive the full chain of lower levels of traceability …". Should this be recognised as correct, as I think it is, the Laboratories could continue to get new values of the constants from their apparatuses, namely of $h$, still after the issue in 2018 of the revised SI, taken them into account until the problem is fixed—and possibly a modified value for $h$ is notified and then adopted.

Authoritative scientists may agree, I think, on the fact that science has no fixed deadlines, outside achievement of facts that bring to a convincing inter-subjective decision of considering the goal sufficiently achieved—i.e., sound and fitting the purpose for the scientists and for the users, in a given moment of science history. Actually, decisions taken "once and for ever" are not typical of science.

As to the SI Brochure, in [CCU 22th] it was stated to be "a CIPM document requested by the CGPM": this means that, should it not adequately informative [Pavese 2017] (the CCU also drafts a CGPM Resolution, then approved by the CIPM, to be approved by the CGPM as the final formal decision [Pavese 2018]), one might wonder how the CGPM can discuss an issue for which no informative document is supplied by the BIPM.

**Appendix – Comment to [Possolo *et al.* 2018] about the results on the numerical value of the Planck constant**

This excellent paper is strongly assertive about the fact that, according to their sound statistical analyses, a sufficient level of confidence has been reached from the available results to proceed to the redefinition of the unit of mass by means of the Planck constant according the present plan to be finalised in Autumn 2018.

However, *their statistical analysis of the experimental data refers to the 2014 CODATA adjustment and related data only*. The 2017 CODATA adjustment—whose preprint was available on 11 September 2017 and is now accepted by the Journal—is only considered in the discussion of the ultimate stability of the recommended numerical value for *h*. However, the scale in Fig. 2 in [Possolo *et al.* 2018] does not allow appreciating the claimed reached stability of $10 \times 10^{-9}$ relative. On the contrary, Fig. 2 (its sub-Section and Table 1) in this manuscript clearly shows that the changes in the available data from 2014 to 2017 are *significant*, especially concerning the results obtained with the Kibble balance—shortly commented in [Possolo *et al.* 2018].

Thus, if left alone, that publication, already available online, might bring to misleading conclusions as it is partially obsolete. Considering the undisputed authority of the authors of that paper, an integration of it with a full analysis of the 2017 data and an updated conclusion seems a necessary supplement.

ArXiv physics.data-an 1512.03668v5